\documentclass{JHEP}

\title{Killing spinors on supersymmetric P-branes}
\author{Rodrigo  Aros \\Departamento de Ciencias F\'isicas \\ Universidad Andr\'es Bello, Sazie 2315, Santiago,
Chile\\E-mail:\email{raros@unab.cl}}
\author{Mauricio Romo\\Department of Physics \\ University of California, Santa Barbara, CA 93106, USA\\E-mail:\email{mromo@physics.ucsb.edu}}


\abstract{A class of $p$-brane solutions for supersymmetric gravity
theories with negative cosmological constant are proposed and
analyzed. The solutions are purely bosonic and contain a worldsheet
and a transverse section. The classification relays on the number of
intrinsic Killing spinors on the worldsheet and the transverse
section. A explicit discussion of the classification is performed
for the four dimensional worldsheet case.}

\begin{document}
\section{Introduction}

The presence of a negative cosmological constant has some very
interesting consequences. For instance it extends the usually called
topological censorship theorem \cite{Woolgar:1999yi} allowing black
holes with topologically non-trivial transverse sections
\cite{Poly}. Analogously, spaces which can be interpreted as
extended objects, namely $p$-branes, are also allowed. These spaces
are interesting since they may be useful as generalizations of the
geometries necessary for Kaluza Klein and Randall Sundrum schemes,
thus they should have a role in dimensional reduction in the
presence of negative cosmological constant, $\Lambda<0$.

In the Kaluza Klein scheme, see appendix \ref{KKsection} for a
review, one considers a vielbein of the form
\[
\tilde{e}^{i} = e^{i}(x)\textrm{ and }\tilde{e}^{m} =
\phi^{m}_{\hspace{1ex} l}(x)(A^{l}(x) + \theta^{l}).
\]
The above geometry is in fact a fiber bundle where, roughly
speaking, $\hat{e}^{i}$ describes the base and $\theta^{l}$ the
fiber. Indeed $\theta^{i}$ stands for a Maurer Cartan basis on the
extra dimensions. $A^{l}$ stands for the gauge fields and
$\phi^{m}_{\hspace{1ex} l}$ for $(d-4)^{2}$ scalars fields.  Here
$i,j=1\ldots p$ and $m,n=p+1 \ldots p+q+1$, where $p$ and $q$ are
the dimension of the base and the transverse section respectively.
The $\{x\}$ stands for generic coordinate system on the observed $p$
dimensions.

One can easily realize that the description of a fiber bundle with a
non vanishing cosmological constant, $\Lambda\neq 0$, should differ
from the $\Lambda=0$ case. In fact, even the spaces to be casted as
backgrounds differ. For $\Lambda=0$ a natural background is a flat
space, or a least Ricci flat one, which implies that the scalar
fields, $\phi^{m}_{\hspace{1ex} l}(x)$, be constant and the
vanishing of every gauge field, $A^{l}$. Conversely, for
$\Lambda\neq 0$ one should expect that the background be constant
curvature manifolds, thus the extra higher dimensions need
\textit{warp} factors.

Aside of the Kaluza Klein construction arose the Randall Sundrum
construction. In this case the entire space has a negative
cosmological constant and unlike Kaluza Klein approach here the
observed universe is located for a particular value of the radial
coordinate $r$.  The spaces considered can be described by $e^{i} =
e^{-\frac{r}{2l}}\hat{e}^{i}(x)$ and $e^{5} = dr$, where
$\hat{e}^{i}$ stands for the observed four dimensions.

In order to incorporate Kaluza Klein and Randall Sundrum schemes
into a single framework with a negative cosmological constant one
can consider
\begin{equation}\label{GeneralizationKKprimer} e^{i} =
e^{-\frac{r}{2l}} \hat{e}^{i}(x),\textrm{   } e^{5}= dr\textrm{  and
} e^{m} = e^{-\frac{r}{2l}}\phi^{m}_{\hspace{1ex} l}(x) (A^{l}(x) +
\theta^{l}).
\end{equation}
The dimension of the space above is given by $d=p+q+1$. It is direct
to prove that this spaces in fact describes a fiber bundle on a
worldsheet as the effective action reads
\[
\int ( R^{(d)} + \Lambda) \sqrt{g^{(d)}} d^{d}x\equiv
\int_{\mathcal{M}_{4}\times \mathbb{R}} \left(\left( R +
\frac{1}{4}g_{ij} F^{i}_{\hspace{1ex} \mu\nu}
F^{j\hspace{1ex}\mu\nu} + \Lambda + \cdots\right)
(\det{\phi^{m}_{\hspace{1ex} j}})\right) e^{-(d-1)\frac{r}{2l}}
\sqrt{g} d^{4}x dr,
\]
where $g_{ij}(x)=\phi^{m}_{\hspace{1ex} i}(x)\phi^{m}_{\hspace{1ex}
j}(x) \delta_{mn}$.

Because $\theta^{i}$ is independent of $x$, the analysis of the
ground state permits to determine the geometry of fiber. In the case
above it is direct to check, as shown in the next sections, that the
ground state of (\ref{GeneralizationKKprimer}) is given by $A^{l}=0$
and the spaces described by $\hat{e}^{a}$ and $\theta^{l}$ should be
at least Ricci flat.  The fundamental point is that the fiber in
(\ref{GeneralizationKKprimer}) must be a Ricci flat manifold.
Fortunately any Calabi Yau manifold is Ricci flat, however not every
Calabi Yau manifold defines a ground state.

In order to generalize the geometry above to include non Ricci flat
fiber one can argue that a space of the form
\begin{equation}\label{GeneralizationKK}
e^{i} = B(r) \hat{e}^{i}(x),\textrm{   } e^{5}= C(r) dr\textrm{  and
}e^{m} = \phi^{m}_{\hspace{1ex} l}(x,r)(A^{l}(x) + \theta^{l}),
\end{equation}
can account for a fiber bundle with a non Ricci flat fiber. In this
case the simplest candidate to be a ground state of this geometry is
given by the line element
\begin{equation}\label{initialgeometry}
ds^{2}=B(r)^{2}(\hat{e}^{i}\hat{e}^{j}\eta
_{ij})+C(r)^{2}dr^{2}+A(r)^{2}( \tilde{e}^{m}\tilde{e}^{n}\eta
_{mn}).
\end{equation}
Here $\hat{e}^{i}$ and $\tilde{e}^{m}$ stands for intrinsic
vielbienen on the world sheet and transverse sections respectively.
Although the final idea is to consider the worldsheet the observed
four dimensional world and the directions in the transverse section
as the fiber still one can discussed a geometry as
(\ref{initialgeometry}) in a general ground, therefore for now the
worldsheet will be considered a $p$-dimensional manifold.

It must be noted that the solution above (\ref{initialgeometry}) is
meaningful only for dimensions $d\geq 5$. In fact one should
consider that the worldsheet and the transverse section at least two
dimensional manifolds in order to have a non trivial case. For $d=5$
the model is suitable for a $U(1)$ fiber.

Before to proceed, a final comment on the spaces described above is
worth to be made to reinforce the aim to analyze them. Since the de
Sitter group has no supersymmetric extension one could have problems
to reconcile supersymmetry with the currently observed positive
cosmological constant. The spaces above solve this problem nicely.
The space above allows to consider a positive cosmological constant
worldsheet, $(\hat{e}^{i})$, in a supersymmetric context because it
is immersed in higher dimensional negative cosmological constant
space. This is actually connected with that any de Sitter space can
be considered a subgroup of higher dimensional anti de Sitter group,
\textit{e.g.}, $SO(d,1)\subset SO(d+1,2)$.

\subsection*{Killing Equation}
The definition of a genuine background can be conceptually
difficult. To address this problem here these spaces will be studied
as solutions of a generic supergravity theory (see \emph{e.g.}
\cite{Salam:1989fm}). In this context a bosonic configuration, a
space in this case, can be casted as a ground state if it is
invariant under supersymmetry transformations. It is also a
candidate to be a BPS state.

Considering a purely gravitational configuration, the arguments
above reduce to determine the spaces where the equation
\begin{equation}
\delta \psi = \nabla \epsilon :=\left( d+A\right) \epsilon=0,
\label{KillingAdS}
\end{equation}
where $A$ is a connection for either Poincar\'e or the anti de
Sitter groups, can be solved. In principle one can solve this
equation also for a connection for de Sitter group, however since
this lacks of a supersymmetric extension this case is not usually
considered. In this work only the anti de Sitter group will be
considered.

It is worth to mention that for some related subjects, as a proof of
the positivity of the energy \cite{Witten:1981mf}, the existence of
a supersymmetric extension is sufficient but not necessary.

In \cite{Aros:2002rk} this approach to identify ground states proved
to be successful for spaces with topologically non-trivial
transverse sections mentioned above \cite{Poly}. In this work an
extension of this idea will be used to classify the spaces that can
represent $p$-brane ground states.

The connection of the anti de Sitter group is given by
\begin{equation}
A=\frac{1}{2}\omega ^{ab}J_{ab}+\frac{1}{l}e^{a}J_{a}, \label{A}
\end{equation}
where $\omega^{ab}$ is a Lorentz connection, $(J_{a},J_{ab})$ are
the generators of AdS group. $a,b=1\ldots d$, with $d$ the dimension
of the space. $l$ is called the AdS radius and is related to the
negative cosmological constant by $\Lambda=-(d-1)(d-2)/(2l^{2})$.

The curvature $F=dA + A \wedge A$ reads
\begin{equation}\label{F}
F=\frac{1}{2}\bar{R}^{ab}J_{ab}+\frac{1}{l} T^{a}J_{a} \textrm{ with
} \bar{R}^{ab} = \left(R^{ab}+\frac{1}{l^{2}}e^{a}\wedge
e^{b}\right),
\end{equation}
where $R^{ab}=d\omega ^{ab}+\omega _{\,\,c}^{a}\wedge\omega ^{cb}$
is the curvature two-form and $T^{a}=de^{a}+\omega^{a}_{\,\,b}
\wedge e^{b}$ is the torsion two form. Although in principle one
could expect that spaces with non vanishing torsion could admits
Killing spinors, none has been found.

\subsection*{Spaces}
The classification of the spaces where Eq.(\ref{KillingAdS}) has
solution arises from the fact that spinors may transform
nontrivially under parallel transport along a closed loop. Indeed
the maximal number of supersymmetries of a Euclidean manifold was
shown to be determined by its holonomy group \cite{Wang1989}, which
are classified by Berger's theorem. This later has been extended to
semi-Riemannian manifolds, and among them Lorentzian manifolds, in
\cite{Bohle2003}. The mathematical construction which allowed the
classification of a simply connected, complete and irreducible
Einstein manifold $X$ of positive scalar curvature \cite{Bar93} (see
also \cite{Figueroa-O'Farrill:1999va}) used the conifold mapping
between $X$ and the cone over $X$, which is a Ricci flat manifold.
The results in \cite{Aros:2002rk, AMT2002} and in the next sections
can be understood as generalizations of that construction. Here it
will be established a correspondence between the Killing spinors of
the whole space and those of the intrinsical geometries of two sub
manifolds which foliate it, a world sheet and a transverse section.

It is worth to mention that the classification of complete,
connected, irreducible Riemannian (Euclidean) manifolds where
Killing spinors exist is well known in the Mathematical literature
\cite{Bar93,Baum721989,Baum731989,BaumKath1999,Baum-Book}. This was
distilled in \cite{Aros:2002rk} to classify the black hole ground
state geometries with topologically non-trivial transverse sections.
Similarly the classification for Lorentzian spaces is also known
\cite{Bohle2003}. Precisely this classification will be used in this
work.

Non simply connected manifolds can be obtained from simply connected
ones by making identification along the orbits of the symmetries
-without fixed points- of the manifold. This in general introduces
noncontractible loops which may further reduce the number of Killing
spinors, and thus the number of supersymmetries. Obviously this must
be studied case by case.

\section{An extended object and a ground state}

The natural vielbein for (\ref{initialgeometry}) is given by
\begin{equation}\label{Vielbein}
e^{i}=B(r)\hat{e}^{i}, \qquad e^{r} =C(r)dr, \qquad
e^{n}=A(r)\tilde{e}^{n}.
\end{equation}
By restricting to torsion free spaces the spin connection,
$\omega^{ab}$, is given by
\begin{eqnarray*}
\omega^{ij} =\hat{\omega}^{ij}, && \omega^{mn} =\tilde{\omega}^{mn},\\
\omega^{ir} =\frac{B(r)^{\prime }}{C(r)B(r)}e^{i}, && \omega^{mr} =\frac{A(r)^{\prime }}{C(r)A(r)}e^{m}, \\
\end{eqnarray*}
where $\hat{\omega}^{ij}$ and $\tilde{\omega}^{mn}$ are the
intrinsic Levi Civita spin connections of the world sheet and
transverse section.

The Riemann curvature reads
\begin{eqnarray}
R^{ij} &=&\hat{R}^{ij}-\left( \frac{\ln (B(r))^{\prime
}}{C(r)}\right)
^{2}e^{i}\wedge e^{j}\nonumber  \\
R^{ir} &=&-\frac{1}{B(r)C(r)}\left( \frac{B(r)^{\prime
}}{C(r)}\right)
^{\prime }e^{i}\wedge e^{r} \nonumber \\
R^{im} &=&-\frac{\ln (B(r))^{\prime }\ln (A(r))^{\prime }}{C(r)^{2}} e^{i}\wedge e^{m} \label{curvatures}\\
R^{rm} &=&-\frac{1}{A(r)C(r)}\left( \frac{A(r)^{\prime }}{C(r)}\right)^{\prime }e^{r}\wedge e^{m}\nonumber  \\
R^{mn} &=&\tilde{R}^{mn}-\left( \frac{\ln (A(r))^{\prime
}}{C(r)}\right) ^{2}e^{m}\wedge e^{n}\nonumber,
\end{eqnarray}
where $\hat{R}^{ij}=\hat{R}^{ij}(\hat{\omega}^{ij})$ and
$\tilde{R}^{mn}=\tilde{R}^{mn}(\tilde{\omega}^{mn})$ are the
intrinsical two forms of curvature of the world sheet and transverse
section respectively.

From now on the language of different forms will be understood, thus
the $\wedge$ product will be omitted.

\section{A constant curvature extended object}

The integrability condition of the Killing spinor equation
(\ref{KillingAdS}) reads
\begin{equation}
\nabla \nabla \epsilon =F\epsilon =0.  \label{Integrability}
\end{equation}
This is trivially satisfied by the vanishing of $T^{a}$ and
$R^{ab}+l^{-2} e^{a} e^{b}$, therefore constant curvature manifolds
are natural candidates to ground sates. However in higher dimensions
$(d\geq4)$ the most general solution is not necessarily a constant
curvature manifold.

Non trivial negative constant curvature solutions
\cite{Aminneborg:1996iz} can be constructed as identifications of
the form AdS/$\Gamma$ where $\Gamma$ is subgroup of AdS, see for
instance \cite{Banados:1998dc}, and by excising regions. This is for
instance the case of the BTZ black hole \cite{Banados:1993gq}.
Because singularities or misbehaved regions must be forbidden in the
case of a ground state $\Gamma$ in that case must be a subgroup of
AdS without fixed points \cite{Wolf1984}. Furthermore this kind of
spaces are solutions of any Lovelock gravity with a single negative
cosmological constant.

From Eq.(\ref{curvatures}) one can determine that in order to
$\bar{R}^{ab}$ vanish then either the world sheet and the transverse
section must be constant curvature manifolds. If $\beta$ is the
curvature of the world sheet and $\alpha$ the curvature of the
transverse section then a generic solution reads
\begin{equation}\label{ABgen}
A(r)^{2}= -\frac{\alpha}{\beta}B(r)^{2}-\alpha \textrm{ and }
C(r)=\sqrt{\frac{(B')^2}{\beta} + \frac{(A')^2}{\alpha}},
\end{equation}
with $B(r)$ arbitrary. The arbitrariness of $B(r)$ is consequence of
that $C(r)$ could be transformed into any function by a redefinition
of the radial coordinate $r$.

For simplicity, without lost of generality, and trying to make
contact with previous known solutions one could take
$\beta=-\alpha=\pm 1$ and restrict to solutions of the form
\begin{equation}\label{ABgen}
B(r)^{2}= l^{2}\left[\frac{r^{2}+C_{1}}{(C_{1}-C_{2})}\right]
\Rightarrow A(r)^{2}= l^{2}\left[\frac{r^{2}+
C_{2}}{(C_{1}-C_{2})}\right],
\end{equation}
where $C_{1},\,C_{2}$ are arbitrary constant. The solution with
$\beta=\alpha=0$ also exists and can be obtained from (\ref{ABgen})
in the limit $\sqrt{C_{1}}\rightarrow \sqrt{C_{2}}$ with a
redefinition of $\beta$ and $\alpha$.

For an explicit example of this kind of geometries see appendix
(\ref{ExplicitConstantCurvature}).

\section{Beyond constant curvature manifolds}

In the previous section was shown that constant curvature solutions,
in the form of Eq.(\ref{initialgeometry}), exist. However one can
readily explore a generalization of the solutions above by
preserving the form of $A(r)$, $B(r)$ and $C(r)$ (see
Eq.(\ref{ABgen})) but leaving the world sheet or transverse section
to be determined by the equations of motion. This is implicit in the
model above. Indeed, these non constant curvature solutions can also
be candidates to ground states. In this case the curvature is merely
given by
\begin{equation}\label{effectiveCurvatures}
\bar{R}^{ab}=\left[
\begin{array}{c|c|c}
\hat{R}^{ij}- \beta \hat{e}^{i}\hat{e}^{j} & 0 & 0 \\
 \hline 0 & 0 & 0 \\
  \hline 0 & 0 & \tilde{R}^{mn}+ \beta \tilde{e}^{m}\tilde{e}^{n}
\end{array}\right],
\end{equation}
with $\beta=\pm 1,0$. Notice that the curvature $\bar{R}^{ab}$ is
completely determined in terms of intrinsic elements of the world
sheet and the transverse section.

\section{Theories to be considered}

As mention above the analysis of the spaces supporting Killing
spinors basically is to determine the ground sate of a gravitational
theory. Nonetheless this only makes sense if there is a supergravity
theory that supports the gravitational theory. Unfortunately, to the
knowledge of these authors, only two of the Lovelock theories have
well established supergravity extensions, Einstein and Chern Simons
gravities. Because of that the analysis will be restricted to these
cases.

The Chern Simons equation of motion ($d=2n+1$) reads
\begin{equation}\label{LovelockEq}
\varepsilon_{b}= \varepsilon_{ba_{1}\ldots a_{2n}} \bar{R}^{a_{1}
a_{2}} \ldots \bar{R}^{a_{2n-1} a_{2n}} =0.
\end{equation}

For $p$ and $q$ even the equation above (\ref{LovelockEq}) reduces,
considering the curvatures (\ref{effectiveCurvatures}), to the
single equation
\begin{equation}\label{EffectiveCSEq}
(\hat{R}^{i_{1}i_{2}}- \beta
\hat{e}^{i_{1}}\hat{e}^{i_{2}})\ldots(\hat{R}^{i_{p-1}i_{p}}- \beta
\hat{e}^{i_{p-1}}\hat{e}^{i_{p}})(\tilde{R}^{l_{1}l_{2}}+ \beta
\tilde{e}^{l_{1}}\tilde{e}^{l_{2}})\ldots(\tilde{R}^{l_{q-1}l_{q}}+
\beta \tilde{e}^{l_{q-1}}\tilde{e}^{l_{q}})\varepsilon_{i_{1}\ldots
i_{p} l_{1}\ldots l_{q}} = 0.
\end{equation}
Since the intrinsic geometries of the worldsheet and transverse
section are completely independent the equation above
(\ref{EffectiveCSEq}) actually separates into the two equations
\begin{equation}\label{Euler}
\begin{array}{l}
(\hat{R}^{i_{1}i_{2}}- \beta
\hat{e}^{i_{1}}\hat{e}^{i_{2}})\ldots(\hat{R}^{i_{p-1}i_{p}}- \beta
\hat{e}^{i_{p-1}}\hat{e}^{i_{p}})\varepsilon_{i_{1}\ldots i_{p}} = 0\textrm{ or},\\
(\tilde{R}^{l_{1}l_{2}}+ \beta
\tilde{e}^{l_{1}}\tilde{e}^{l_{2}})\ldots(\tilde{R}^{l_{q-1}l_{q}}+
\beta \tilde{e}^{l_{q-1}}\tilde{e}^{l_{q}})\varepsilon_{l_{1}\ldots
l_{q}} = 0.
\end{array}
\end{equation}
Therefore to have a solution of Eq.(\ref{EffectiveCSEq}) is enough
that either the worldsheet or the transverse section have vanishing
Eq.(\ref{Euler}). Unfortunately this leaves respectively either the
transverse section or the world sheet undetermined. A trivial
solution is therefore that either the transverse section or the
worldsheet be constant curvature manifolds.

Finally, for $p$ and $q$ odd, the other case of Chern Simons
gravity, the result is rather trivial since the equation
(\ref{LovelockEq}) is identically satisfied leaving both worldsheet
and transverse section undetermined. These results only confirm that
Chern Simons theories are deeply complex for they have many
sub-sectors with high degrees of degeneracy. This is a well known
situation and has proven to be major obstacle to achieve a
perturbative analysis of the Chern Simons gravity.

The Einstein-Hilbert case is straightforward and far more
restrictive. In any dimension, considering
(\ref{effectiveCurvatures}), Einstein equations of motion decouple
into the two set of Einstein equations
\[
(\hat{R}^{i_{1}i_{2}}- \beta
\hat{e}^{i_{1}}\hat{e}^{i_{2}})\hat{e}^{i_{3}}\ldots\hat{e}^{i_{p-1}}
\varepsilon_{i_{1}\ldots i_{p}}=0 \textrm{ and }
(\tilde{R}^{l_{1}l_{2}}+ \beta
\tilde{e}^{l_{1}}\tilde{e}^{l_{2}})\tilde{e}^{l_{3}}\ldots\tilde{e}^{l_{q-1}}
\varepsilon_{l_{1}\ldots l_{q}}=0,
\]
determining that the worldsheet and the transverse section satisfy
Einstein equations with a cosmological constant $\pm \beta$ on their
own.

\section{Killing spinors and representations}

Returning to the Killing spinor equation (\ref{KillingAdS}). By
expressing generators in terms of $d$ dimensional Dirac matrices as
$J_{ab}=\frac{1}{2}\Gamma _{ab}$ and $J_{a}=\frac{1}{2}\Gamma _{a}$
the connection one-form $A$ reads
\[
A =\left(\frac{1}{2l}C(r)\Gamma _{1}\right) +
\frac{1}{2}\left(\frac{B(r)}{l}\hat{e}^{i} \Gamma _{i} +
\hat{\omega}^{ij}\Gamma _{i}\Gamma _{j}\right) +
\frac{1}{2}\left(\frac{A(r)}{l}\tilde{e}^{m} \Gamma _{m} +
\tilde{\omega}^{mn}\Gamma _{m}\Gamma _{n}\right).
\]
Formally the solution of the Killing spinor equation (\ref
{KillingAdS}) reads
\begin{equation}
\epsilon =e^{-\Gamma_{1}H(r)}\eta, \label{Solution}
\end{equation}
where
\[
H(r) = \frac{1}{2l}\int C(r) dr
\]
and $\eta$ satisfies the equation
\begin{equation}\label{KillingTransverse}
\left( d+\hat{A}+\tilde{A}\right) \eta =0
\end{equation}
and
\begin{equation}\label{connections}
\hat{A} =  \frac{1}{2}\hat{\omega}^{ij}J_{ij}+\hat{e}^{i}P_{i},%
\qquad%
\tilde{A} = \frac{1}{2}\tilde{\omega}^{mn}J_{mn}+\tilde{e}^{m}P_{m},
\end{equation}
where
\begin{equation}\label{Ps}
P_{i} =\frac{1}{2}(P_{-}-\beta P_{+})\Gamma_{i},%
\qquad%
P_{m} =\frac{1}{2}\left(P_{-}+\beta P_{+}\right) \Gamma _{m},
\end{equation}
with $P_{\pm }:=\frac{1}{2}(1\pm \Gamma _{1})$.

Note that since $[P_{i},P_{j}]=-\beta J_{ij}$, and
$[P_{m},P_{n}]=\beta J_{mn}$, therefore the sets
$\left\{P_{n},J_{mn}\right\}$ and the sets $\left\{P_{n},J_{mn}
\right\}$ form respectvely \emph{reducible} representation for
$SO(d-1,1)$, $SO(d-2,2)$, or $ISO(d-1,1)$ depending on whether
$\beta =1,-1,$ or $0$, respectively.

The case $\beta=0$ decouples in the following form; Let
$\eta_±=P_±\eta$ which separates the Eq.(\ref{KillingTransverse})
becomes
\begin{equation}
d\eta
_{+}+\frac{1}{4}\left(\hat{\omega}^{ij}\Gamma_{ij}+\tilde{\omega}^{mn}\Gamma
_{mn} \right)\eta _{+}=0\, \label{FlatEquation1}
\end{equation}
and
\begin{equation}
d\eta _{-}+\frac{1}{4}\left(\hat{\omega}^{ij}\Gamma
_{ij}+\tilde{\omega}^{mn}\Gamma _{mn} \right)\eta _{-}=\frac{1}{2}
\left(\Gamma _{i}\hat{e}^{i}+\Gamma _{m}\tilde{e}^{m}\right)\eta
_{+}. \label{FlatEquation2}
\end{equation}
Flat spaces unfortunately do not have a natural scale to define,
unlike constant curvature manifold where the cosmological constant
defines a scale. To introduce one scale one can wrap one direction
in the manifold yielding spaces of the form
\begin{equation}\label{wrapped}
 ds^2 = d\phi^ 2 + d\Sigma_{0},
\end{equation}
where $\phi$ defines a circle and $\Sigma_{0}$ is also a $\beta=0$
submanifold. The presence of this cycle determines that $\eta_{+}=0$
and therefore that $\eta_{-}$ (\ref{FlatEquation2}) satisfies an
equation for the Lorentz group.

Nonetheless one can still consider the general case. In principle
the solution of $\eta _{-}$ can be written in terms of $\eta _{+}$,
which in turn satisfies an equation for the Lorentz group. The
consistency condition for Eq.(\ref{FlatEquation2}) gives the same
information as Eq.(\ref{FlatEquation1}), \emph{i.e.},
\[
\left(\hat{R}^{ij}\Gamma_{ij}+\tilde{R}^{mn}\Gamma _{mn} \right)\eta
_±=0
\]

\subsection*{Representations}

Let us separate the cases according to the dimension of world sheet
and transverse section. Recalling that $p$ and $q$ are the
dimensions of the worldsheet and the transverse section respectively
one can propose the following three representation according to $q$
and $p$.

\begin{itemize}
\item $\mathbf{p=2m}$ \textbf{and }$\mathbf{q=2n}$.

In this case the dimension of the space is $d=2(m+n)+1$, thus the
dimension of the spinor $\eta$ is $2^{m+n}$. This allows to propose
a representation where the spinor can be written as
\[
  \eta = \hat{\eta}\otimes\tilde{\eta}
\]
where $\hat{\eta}$ and $\tilde{\eta}$ are genuine spinor on the
worldsheet and the transverse section, with dimension $2^{m}$ and
$2^{n}$ respectively. The representation of the $\Gamma$ matrices is
given by
\[
\Gamma_{i} = \gamma_{i}\otimes M_{\beta},\qquad
\Gamma_{1}=\gamma\otimes\sigma \textrm{ and
}\Gamma_{m}=N_{\beta}\otimes\sigma_{m}
\]
where $N_{1}=I_{2^{m}}\otimes I_{2^{n}}$,
$N_{-1}=-i\gamma\otimes\sigma$ and $M_{1}=i\gamma\otimes\sigma$,
$M_{-1}=I_{2^{m}}\otimes I_{2^{n}}$. $\gamma$ and $\sigma$ are the
proportional to $\gamma_{2m+1}$ and $\sigma_{2n+1}$ and satisfy
$\gamma^2=I$ and $\sigma^2=I$ respectively.

This representation, depending on $\beta$ yields the connections
Eq.(\ref{connections})

\begin{enumerate}
\item $\beta=-1$
\[
\hat{A} = \left(\frac{1}{2}\hat{\omega}^{ij}\gamma_{i}\gamma_{j}
+\frac{1}{2}\hat{e}^{i}\gamma_{i}\right)\otimes I_{2^{n}}, \qquad
\tilde{A}
=I_{2^{m}}\otimes\left(\frac{1}{2}\tilde{\omega}^{mn}\sigma_{m}\sigma_{n}+\frac{i}{2}\tilde{e}^{m}\sigma_{m}\right)
\]
\item $\beta=1$
\[
\hat{A} = \left(\frac{1}{2}\hat{\omega}^{ij}\gamma_{i}\gamma_{j}
+\frac{i}{2}\hat{e}^{i}\gamma_{i}\right)\otimes I_{2^{n}}%
\qquad%
\tilde{A} =
I_{2^{m}}\otimes\left(\frac{1}{2}\tilde{\omega}^{mn}\sigma_{m}\sigma_{n}+\frac{1}{2}\tilde{e}^{m}\sigma_{m}\right)
\]
\item $\beta=0$
This case subtly different. Recalling the projection in terms of
$P_±$ one obtains
\[
\eta_{+} = \hat{\eta}_{+}\otimes\tilde{\eta}_{+}+ \hat{\eta}_{-}\otimes\tilde{\eta}_{-},%
\textrm{ and }%
\eta_{-} = \hat{\eta}_{+}\otimes\tilde{\eta}_{-}+
\hat{\eta}_{+}\otimes\tilde{\eta}_{-}
\]
where $\gamma \hat{\eta}_±=\pm\hat{\eta}_±$ and $\sigma
\tilde{\eta}_±=\pm\tilde{\eta}_±$. Restricting as mentioned before
spaces with a wrapped direction of the form (\ref{wrapped}), which
determines $\eta_{+}=0$, and the representation
\[
\Gamma_{i} = \gamma_{i}\otimes I_{2^{n}},\qquad
\Gamma_{1}=\gamma\otimes\sigma \textrm{ and
}\Gamma_{m}=\gamma\otimes\sigma_{m}
\]
one can demonstrate that Eqs. (\ref{FlatEquation1},
\ref{FlatEquation2}) become a Killing equation on the world sheet
and transverse section provided that $\hat{\eta} _{+}=\tilde{\eta}
_{+}=0$ or $\hat{\eta} _{-}=\tilde{\eta} _{-}=0$
\begin{equation}
\left(d+\frac{1}{4}\hat{\omega}^{ij}\gamma _{ij}\right)\hat{\eta}
_±=0
\end{equation}
\begin{equation}
\left(d+\frac{1}{4}\tilde{\omega}^{mn}\sigma
_{mn}\right)\tilde{\eta} _±=0
\end{equation}

\end{enumerate}

This proves that the representations above indeed separates the
Killing spinors equation (\ref{KillingAdS}) into worldsheet and
transverse section. Therefore this proves that the problem of
Killing spinors on the space has been reduced to the Killing spinor
problem on the worldsheet and the transverse section.

\item $\mathbf{p=2m+1}$ \textbf{and }$\mathbf{q=2n}$.

In this case the dimension of the space is $d=2(m+n)+2$, thus the
dimension of the spinor $\eta$ is $2^{m+n+1}$. This allows to
propose a representation where the spinor can be written as
\[
  \eta = \hat{\eta}\otimes\tilde{\eta}\otimes\bar{\eta}
\]
where $\hat{\eta}$ and $\tilde{\eta}$ are genuine spinor on the
worldsheet and the transverse section, with dimension $2^{m}$ and
$2^{n}$ respectively. $\bar{\eta}$ is a two dimensional constant
spinor.

The representation of the $\Gamma$ matrices is given by
\[
\Gamma_{i} = \gamma_{i}\otimes I_{2n}\otimes \sigma_{x},\qquad
\Gamma_{1}=I_{2m}\otimes I_{2n} \otimes \sigma_{z} \textrm{ and
}\Gamma_{m}=I_{2m}\otimes\sigma_{m}\otimes\sigma_{y}
\]
 $I_{2m,2n}$ are the
identity matrices in $2^{m}$ and $2^{n}$ dimensions.

In this representation, provided $\sigma_{y}\bar{\eta}=\pm
\bar{\eta}$ when $\beta=1$ and $\sigma_{x}\bar{\eta}=\pm \bar{\eta}$
when $\beta=-1$, the connection Eq.(\ref{connections}) splits as

\begin{enumerate}
\item $\beta=-1$
\[
A\eta =
\left(\frac{1}{2}\hat{\omega}^{ij}\gamma_{i}\gamma_{j}\otimes
\pm\frac{1}{2}\hat{e}^{i}\gamma_{i}\right)\hat{\eta}\otimes
\tilde{\eta}\otimes\bar{\eta}
+\hat{\eta}\otimes\left(\frac{1}{2}\tilde{\omega}^{mn}\sigma_{m}\sigma_{n}\pm\frac{i}{2}
\tilde{e}^{m}\sigma_{m}\right)\tilde{\eta}\otimes\bar{\eta}
\]
\item $\beta=1$
\[
A\eta = \left(\frac{1}{2}\hat{\omega}^{ij}\gamma_{i}\gamma_{j} \mp
\frac{i}{2}\hat{e}^{i}\gamma_{i}\right)\hat{\eta}\otimes\tilde{\eta}\otimes\bar{\eta}
+
\hat{\eta}\otimes\left(\frac{1}{2}\tilde{\omega}^{mn}\sigma_{m}\sigma_{n}\pm\frac{1}{2}\tilde{e}^{m}\sigma_{m}\right)\tilde{\eta}\otimes\bar{\eta}
\]
\item $\beta=0$
In this case using the projection in terms of $P_±$ one obtains
\[
\eta_± = \hat{\eta}\otimes\tilde{\eta}\otimes\bar{\eta}_±
\]
where $\sigma_{z} \bar{\eta}_±=\pm\bar{\eta}_±$. One can demonstrate
that Eqs. (\ref{FlatEquation1}\ref{FlatEquation2}) splits on the
following Killing equations on the world sheet and transverse
section provided $\bar{\eta}_{+}=0$
\begin{equation}
\left(d+\frac{1}{4}\hat{\omega}^{ij}\gamma _{ij}\right)\hat{\eta} =0
\end{equation}
\begin{equation}
\left(d+\frac{1}{4}\tilde{\omega}^{mn}\sigma
_{mn}\right)\tilde{\eta} =0
\end{equation}

\end{enumerate}

As previously in this case representation also separates the Killing
spinors equation (\ref{KillingAdS}) into worldsheet and transverse
section. Once again the problem of Killing spinors on the space has
been reduced to find Killing spinors on the worldsheet and the
transverse section.

\item $\mathbf{p=2m+1}$ \textbf{and }$\mathbf{q=2n+1}$.

This case is totally analogous to the $p=2m+1$, $q=2n$ case analyzed
before.\\
\end{itemize}

\section{Lorentzian manifolds}\label{sectionClassification}

In this section is summarized the classification of Lorentzian
manifolds allowing Killing spinors, therefore it can be skipped for
those well familiarized with the subject.

In the sections above was shown that the $d$ dimensional Killing
spinor equation reduces to effective equation on the world sheet and
the transverse sections, spaces which can be either Lorentzian or
Euclidean. For simplicity one can consider only effective equations.
Let $\Sigma$ be that manifold where the Killing equation takes the
form
\begin{equation}\label{formalEquation}
\left( d + \frac{1}{2} \omega^{AB}\gamma_{A}\gamma_{B} +
\left(\frac{\sqrt{-\kappa}}{2}\right) e^{A}\gamma_{A}\right)\zeta
\end{equation}
where $\kappa=0,\pm 1$, $\gamma^{A}$ are the corresponding Dirac
matrices and  $\zeta$ is a spinor. It is direct to demonstrate that
$sgn(R(\Sigma))=sgn(\kappa)$ and $R(\Sigma)=0$ if $\kappa=0$.

According to the value of $\kappa$ the $\Sigma$ spaces are given by
\begin{enumerate}
\item $\kappa=0$

If the signature of $\Sigma$ is $(t,s)$ and $\Sigma$ is an
irreducible, simply connected and totally symmetric
pseudo-Riemannian manifold then it has $N$ Killing spinors if and
only if its holonomy group $H$ is on the table below \cite{BaumKath1999}. \\
\begin{center}
\begin{tabular}{||l|c|c|c||}
\hline
$H$ & $t$ & $s$ & $N$\\ \hline\hline%
$SU(a,b)$ & $2a$ & $2b$ & $2$ \\ \hline%
$Sp(a,b)$ & $4a$ & $4b$ & $a+b+1$ \\ \hline%
$G_{2}$ & $0$ & $7$ & $1$ \\ \hline%
$G^{*}_{2(2)}$ & $4$ & $3$ & $1$ \\ \hline%
$G^{\mathbb{C}}_{2}$ & $7$ & $7$ & $2$ \\ \hline%
$Spin(7)$ & $0$ & $8$ & $1$ \\ \hline%
$Spin^{+}(4,3)$ & $4$ & $4$ & $1$ \\ \hline%
$Spin(7)^{\mathbb{C}}$ & $8$ & $8$ & $1$ \\
\hline\hline
\end{tabular}
\end{center}

Note that in the classification above there are no Lorentzian
manifolds. Although this seems rather restrictive actually is only
due to the $\kappa=0$ Lorentzian manifolds that admits Killing
spinors are reducible, for instance Minkowski.

\item $\kappa=-1$

The analysis, in this case, is made in function of the cone $C$ over
$\Sigma$. $C$ is defined as $-dt^{2}+t^{2}d\Sigma^{2}$ and $\Sigma$
has Killing spinors with $\kappa=-1$ if and only if $C$ has Killing
spinors with $\kappa=0$.\\
If $C$ is irreducible the only possible $\Sigma$ is a Lorentzian
Einstein Sasaki manifold with $dim(\Sigma)$ odd.\\
A most comprehensive clasification has been done in \cite{Leitner},
in function of the Dirac current
$V^{A}_{\zeta}=\overline{\zeta}\gamma^{A}\zeta$\\\\
\textbf{Theorem} \textit{Let $\Sigma$ be a Lorentzian manifold with
Killing spinors with $\kappa=-1$\\\\
1. If $\Sigma$ is not Einstein then $\Sigma$ is locally conformally
equivalent to a Brinkmann space with Killing spinor with
$\kappa=0$.\\\\
2. If $V^{A}_{\zeta}V_{A\zeta}$ is constant then\\
i) $V^{A}_{\zeta}V_{A\zeta}=0$ and $\Sigma$ is locally conformally
equivalent to a Brinkmann space with Killing spinor with
$\kappa=0$.\\
ii) $V^{A}_{\zeta}V_{A\zeta}<0$ and $\Sigma$ is a Lorentzian
Einstein Sasaki manifold.\\\\
3. If $C$ is indecomposable and $V^{A}_{\zeta}V_{A\zeta}<0$ then\\
i) $\Sigma$ is locally conformally equivalent to a Brinkmann space
with Killing spinor with $\kappa=0$.\\
ii)$\Sigma$ admits locally a warped product structure of the form
\[
dt^{2}+f(t)^2ds^{2}_{\mathcal{F}}
\]
where $ds^{2}_{\mathcal{F}}$ is the line element of a Lorentzian
Einstein manifold of the table below}

\begin{center}
\begin{tabular}{||l|c||}
\hline
$\mathcal{F}$ & $f(t)$ \\ \hline\hline%
Lorentzian Manifold with Killing spinors and $\kappa=-1$ & $\cosh(t)$ \\ \hline%
Lorentzian Manifold with Killing spinors and $\kappa=0$ & $e^{t}$ \\ \hline %
Lorentzian Manifold with Killing spinors and $\kappa=1$ & $\sinh(t)$ \\
\hline\hline
\end{tabular}
\end{center}

\textit{iii) $\Sigma$ is a Lorentzian Einstein Sasaki manifold (case
$C$
irreducible).\\\\
4. If $V^{A}_{\zeta}V_{A\zeta}< 0$ changes from $0$ to a negative
value then the region $\Omega\subset\Sigma$ where
$V^{A}_{\zeta}V_{A\zeta}=0$ is a hypersurface and $\Sigma \ \Omega$
admits locally a warped product structure as in 3. ii).\\
If the metric does not belong to the cases listed in 3. then
$V^{A}_{\zeta}V_{A\zeta}< 0$ changes from $0$ to a negative value or
there exists a parallel \textrm{2}-form which vanishes in a
Riemannian subspace of $C$ and is a pseudo-K\"ahler form on the
complement.}

\item $\kappa=-1$

Here $\Sigma$ support solutions of Eq.(\ref{formalEquation})
provided it is locally described by the line element
\[
ds^2 = \sigma^2 ds^2_{\mathcal{F}} + \varepsilon dt^2,
\]
where $\varepsilon$ is described by the table below and
$ds^2_{\mathcal{F}}$ is the line element
of the space $\mathcal{F}$ also described in this table; \\
\begin{center}
\begin{tabular}{||l|c|r||}
\hline
$\mathcal{F}$ & $\sigma$ & $\varepsilon$ \\ \hline\hline%
Riemannian Manifold with Killing spinors and $\kappa=1$ & $\cosh(t)$ & $-1$ \\ \hline%
Riemannian Manifold with Killing spinors and $\kappa=0$ & $e^{t}$ & $-1$ \\ \hline %
Riemannian Manifold with Killing spinors and $\kappa=-1$ & $\sinh(t)$ & $-1$ \\ \hline%
Lorentzian Manifold with Killing spinors and $\kappa=1$ & $\cos(t)$ & $1$ \\
\hline\hline
\end{tabular}
\end{center}
\end{enumerate}

\section{Classification of ground states with $p=4$}\label{Classification}
Using the classification above, and that in \cite{Aros:2002rk}, one
can classify the spaces in terms of $p$ and $q$. The analysis
performed in this section is limited to $p=4$ trying to make contact
with the observed four dimensions such that the worldsheet could be
considered as the visible world.

\subsection{$\beta=1$}
In this case most general four dimensional worldsheet has locally
the form of the warped product
\[
ds^2 = \sigma^2 ds^2_{\mathcal{F}} + \varepsilon dt^2.
\]
The classification for $\varepsilon=-1$ is given by the following
table
\begin{center}
\begin{tabular}{||c|r|c||}
\hline
$\sigma$ & $R_{\mathcal{F}}$ & $\mathcal{F}$\\ \hline\hline%
$\cosh(t)$ & $1$ & $S^{3}$, $\mathbb{RP}^{3}$  \\ \hline%
$e^{t}$ & $0$ & $\mathbb{R}^{3}$, $S^{1}\times \mathbb{R}^{2}$, $S^{1}\times S^{1}\times \mathbb{R}$, $S^{1}\times S^{1}\times S^{1}$ \\ \hline%
$\sinh(t)$ & $-1$ & $H^{3}$, $H^{3}/\Gamma$  \\
\hline\hline
\end{tabular}
\end{center}
where $\Gamma$ is a normal subgroup of $H^{3}$ without fixed points
and such that quotient manifold be non-compact.

For $\varepsilon=1$ $\sigma=\cos(t)$ and $\mathcal{F}$ is a three
dimensional Lorentzian manifold which is classified analogously to
the table above by lowing the dimension from 3 to 2. The
only exception is $\mathbb{RP}^{2}$ which is non-orientable so it must excluded from the sub classification.\\

\subsection{$\beta=-1$}

In this case one has a general family of geometries described by the
stationary line element
\begin{equation}
ds^{2}=-N(r)^{2}dt^{2}+\frac{1}{N(r)^{2}}dr^{2}+r^{2}d\sigma^{2}_{\chi}
\end{equation}
where $d\sigma^{2}_{\chi}$ is the line element of a two dimensional
manifold with constant curvature $\chi$ listed in the table below
\begin{center}
\begin{tabular}{||r|c||}
\hline
$\chi$ & $\sigma_{\chi}$\\ \hline\hline%
$1$ & $S^{2}$  \\ \hline%
$0$ & $S^{1}\times \mathbb{R}$, $\mathbb{R}^{2}$, $S^{1}\times S^{1}$ \\ \hline%
$-1$ & $H^{2}$, $H^{2}/\Gamma$ \\
\hline\hline
\end{tabular}
\end{center}
where $\Gamma$ is a normal subgroup of $H^{2}$ without fixed points
and such that quotient manifold be non-compact.

The rest of the spaces have non static worldsheet. One remarkable
example of these spaces \cite{Bohle2003} is described by the line
element
\begin{equation}\label{bohle}
   ds^2 = e^{2u}\left( dx^{2} + f(x,s) ds^2 - 2ds dt \right) + du^2.
\end{equation}
This is not an Einstein space unless $f(x,s)$ be a harmonic function
on $x$ for all $s$, \textit{i.e.}, $f(x,s)=f_{1}(s)x+f_{2}(s)$.

Nonetheless in the general case the space above has a single Killing
spinor given by
\begin{equation}\label{KillingBohle}
  \eta=\frac{e^{-\gamma_{1}\frac{u}{2}}}{f(x,s)^{\frac{1}{4}}}\eta_{0}
\end{equation}
where $\eta_{0}$ is a constant spinor that satisfies
$(\gamma_{0}+\gamma_{3})\eta_{0}=0$ and $(1+\gamma_{1})\eta_{0}=0$.

This space above (\ref{bohle}) admits a compactification along
$\partial_{x}$ which preserves the Killing spinor
(\ref{KillingBohle}) provided $f(x,s)=f(s)$. In this case the space
is an Einstein space though.

\subsection{$\beta=0$}

As mentioned in the previous sections, for  $\beta=0$ there are no
irreducible Lorentzian spaces having Killing spinors. This is due to
the such spaces can always be constructed as a direct product of
spaces of the form
\begin{equation}
   ds^2 = ds_{\mathcal{F}}^{2}-dt^2,
\end{equation}
where $\mathcal{F}$ is a Riemannian manifold with Killing spinors
with $\kappa=0$. Note that this decomposition is due to the space
above is a Ricci flat manifold. Since $\mathcal{F}$ is a three
dimensional manifold with $\kappa=0$ thus it is $\mathbb{R}^{3}$,
$\mathbb{R}^2\times S^1,\mathbb{R}\times(S^1)^2, (S^1)^3$.

\subsection{Possible transverse sections}\label{finalTable}

So far the possible four dimensional worldsheets have been
identified. The classification of the possible transverse sections
is summarized in the following table:
\begin{center}
\begin{tabular}{||c|c|c|c|c||}
\hline
$d$ & $q$ & $\beta=1$ & $\beta=0$ & $\beta=-1$ \\ \hline\hline%
$7$ & $2$ &  $H^{2}$,$H^{2}/\Gamma$ & $\mathbb{R}^2$,
$\mathbb{R}\times S^1$ or $(S^1)^2$
             & $S^{2}$\\ \hline%
$8$ & $3$ & $H^{2}$,$H^{3}/\Gamma$ & $\mathbb{R}^{3}$,
$\mathbb{R}^2\times S^1,\ldots, (S^1)^3$
                           & $S^{3}$, $\mathbb{RP}^{3}$  \\ \hline%
$9$ & $4$ & $H^{4}$,$H^{4}/\Gamma$ & $\mathbb{R}^4$, $\mathbb{R}^3\times S^1,\ldots,(S^1)^4$ & $S^{4}$ \\ \hline%
$10$ & $5$ & \begin{tabular}{c}
              $H^{5}$,$H^{5}/\Gamma$,\\$\frac{1}{z^{2}}(dz^{2}+h_{ij}dx^{i}dx^{j})$

            \end{tabular} & $\mathbb{R}^5$, $\mathbb{R}^4\times S^1,\ldots,(S^1)^5$  &  \begin{tabular}{l}
                                                                               $S^{5}$,$\mathbb{RP}^{5}$,\\
                                                                               Sasaki-Einstein \\
                                                                             \end{tabular}
              \\ \hline%
$11$ & $6$ & $H^{6}$,$H^{6}/\Gamma$  & $\mathbb{R}^6$,
$\mathbb{R}^5\times S^1,\ldots,(S^1)^6$ &
\begin{tabular}{l}
            $S^{6}$, nearly\\
            K\"ahler manifold \\

            \end{tabular} \\
\hline\hline \end{tabular}
\end{center}
where in the third column $\Gamma$ stands for a normal subgroup of
$H^{n}$ without fixed points and such that $H^{n}/\Gamma$ be non
compact. $h_{ij}$ stands for the metric of a Hyperk\"ahler with
holonomy $Sp(2)$ or a Calabi-Yau with holonomy $SU(2)$ manifold.

\section{Conclusions and prospects}

We have classified the families of spaces with 4-brane worldsheets
that support Killing spinors and thus can be casted as genuine
ground states. The classification above can be extended to any
higher dimensions of the worldsheet by a careful reading of the
tables in section \ref{sectionClassification}.

The analysis of the four dimensional case has some interesting
features. The spaces differs depending on the theory considered
though.
\begin{description}
  \item[Einstein theory] In this case the equations of motion force both worldsheet and transverse section be
Einstein manifolds on their own. Therefore, among the spaces
permitted, four dimensional worldsheets and tranverse sections, one
must simply exclude the non Einstein spaces to complete the
classification. This for instance forbids a worldsheet of the form
(\ref{bohle}).

\item[Chern Simons theory] For $d<9$ every worldsheet in section \ref{Classification}
is permitted because the transverse section is a constant curvature
manifold. This is particularly relevant by the presence for
$\beta=-1$ of non-Einstein manifolds.

On the other hand, for $\beta=0,1$ any worldsheet in section
\ref{Classification} is permitted since every possible transverse
section is either a flat or a positive curvature manifold, thus the
entire space is a trivial solution of the Chern Simons equations.

For $\beta=-1$ and $d=9,11$, nonetheless, one must proceed with a
case by case analysis to check if those spaces are solutions of
Chern Simons theory. However there are fundamental examples that are
worth to mention. Recalling the Chern Simons equations reduce to the
multiplication of two independent equations of motion. The equation
of motion for the worldsheet reads
\[
\mathcal{E} = (\hat{R}^{i_{1}i_{2}} \pm
\hat{e}^{i_{1}}\hat{e}^{i_{2}})(\hat{R}^{i_{3}i_{4}}\pm
\hat{e}^{i_{3}}\hat{e}^{i_{4}})\varepsilon_{i_{1}\ldots i_{4}}.
\]
Remarkably the Bohle space in Eq.(\ref{bohle}) solves this equation,
and thus for any transverse section in section \ref{finalTable}, in
$d=9,11$, there is a ground state.
\end{description}

The possible directions to continue this work are many. However, to
proceed to investigate solutions over this ground states seems a
natural next step. For this the presence of Calabi-Yau geometries is
most relevant, in particular for the search of effective gauge
theories on the worldsheet \cite{ArosRomo2007-2}.

Finally it must be stressed that in the context of this work, de
Sitter spaces, in particular four dimensional ones, were naturally
incorporated into a supersymmetric framework. This could be most
relevant to take in the current astrophysical observations and
supergravity into a single unified context.

\appendix
\section{Appendix}
\subsection{Dimensional reduction}\label{KKsection}

To understand the kind of space to be discussed one can review
dimensional reduction. In order to make contact with the rest of
this work compactification in terms of vielbeinen and spin
connections, see for instance \cite{ArosRomoZamorano2006RMF}, will
be discussed. First the presence of those higher dimensions should
generate the arise of non abelian gauge theories in four dimensions.
For this however the additional dimensions can not arbitrary. In
particular one must consider $\mathcal{M}_{d}=\mathcal{M}_{4}\times
G_{d-4}$, being $G_{d-4}$ a group manifold $\mathbf{G}$ or
$\mathbf{G}/\mathbf{H}$ with $\mathbf{H}$ a normal subgroup of
$\mathbf{G}$ \cite{yvonne}. For instance, the spheres
$S^{n}=SO(n+1)/SO(n)$ and $S^{2n-1}=SU(n)/SU(n-1)$ are perfect
candidates for $G_{d-4}$.

The ansatz for the vielbein is $\tilde{e}^{A}$, with $A=0\ldots d$,
\begin{equation}\label{vielbeinDdimensinal}
\tilde{e}^{a} = e^{a}(x)\textrm{ and }\tilde{e}^{m} =
\phi^{m}_{\hspace{1ex} i}(x)(A^{i}(x) + \theta^{i})
\end{equation}
where $\theta^{i}$ is a Maurer Cartan basis on $G_{d-4}$. From a
geometrical point of view $\phi^{m}_{\hspace{1ex i}}$ diagonalizes
the direction along the fiber,\textit{ i.e.}, it satisfies
\[
\phi^{m}_{\hspace{1ex} i}\phi^{m}_{\hspace{1ex} j} \delta_{mn} =
g_{ij}(x) \textrm{ and } g^ {ij}\phi^{m}_{\hspace{1ex}
i}\phi^{n}_{\hspace{1ex} j} = \delta_{mn}.
\]
In four dimensions these fields actually correspond to a collection
of $(d-4)^{2}$ scalars fields.

A spin connection compatible with the symmetries of $\mathbf{G}$ is
given by $\tilde{\omega}^{ab} = \omega^{ab}(x)  +
\psi^{ab}_{\hspace{2ex}i}(x) \theta^{i}$, $\tilde{\omega}^{an} =
\omega^{an}(x)  + \hat{\omega}^{an}_{\hspace{2ex}j}(x)\theta^{j}$
and $\tilde{\omega}^{mn} = \omega^{mn}(x)  +
\hat{\omega}^{mn}_{\hspace{2ex}j}(x)\theta^{j}$. Here
$\psi^{ab}_{i}$ and $\hat{\omega}^{an}_{j}$ are scalars and
$\omega^{ab}$ and $\hat{\omega}^{an}$ are one-form respectively on
$\mathcal{M}_{4}$.

The torsion free equation in $d$ dimensions, $\tilde{T}^{A}=0$,
determines the effective torsion in four dimensions $T^{a} =-
\omega^{a}_{\hspace{1ex} m} \phi^{m}_{\hspace{1ex} j} A^{j}$ where
\[
\omega^{a m} = -E^{a\mu}  \left(\phi^{m}_{\hspace{1ex}
j}\frac{1}{2}F^{j}_{\,\,\mu\nu}dx^{\nu}  +
\partial_{\mu}(g_{ij})\phi^{m\,j} A^{i}\right),
\]
with $F^{i} = dA^{i} + \frac{1}{2}C^{i}_{jk} A^{j}A^{k}$. One can
show that the contorsion is given by $K^{ab} =- g_{ij}
F^{j\,ab}A^{i}$. Torsion does not vanish unless $A^{i}$ be pure
gauge. Indeed torsion can be understood completely in terms of gauge
fields.

This construction will determine the effective theory in four
dimensions. For instance the Einstein Hilbert action in $d$
dimensions is reduced to
\begin{equation}\label{EffectiveEHElecD}
I_{EH}=I_{eff}=\int_{M_{d}} \left( R + \frac{1}{4}  g_{ij}
F^{i}_{\mu\nu} F^{j\,\mu\nu} + \dots\right)
\det({\phi^{m}_{\hspace{1ex} j}})\sqrt{g} d^4x\,
\varepsilon_{i_{1}\ldots i_{d-4}} \theta^{i_{1}}\ldots
\theta^{i_{d-4}},
\end{equation}
where $g_{ij}(x)=\phi^{m}_{\hspace{1ex} i}(x)\phi^{m}_{\hspace{1ex}
j}(x) \delta_{mn}$. Here $R$ is the standard torsion free four
dimensional Ricci scalar. The dots account for derivatives of
$g_{ij}$.

Since this action (\ref{EffectiveEHElecD}) is independent of the
coordinates on $G_{d-4}$ one can integrate them out, thus
\begin{equation}\label{EffectiveEHElecIID}
\hat{I}_{eff}  =\int_{\mathcal{M}_{4}} \left( R + \frac{1}{4}g_{ij}
F^{i}_{\hspace{1ex} \mu\nu} F^{j\hspace{1ex}\mu\nu}+ \ldots\right)
\det({\phi^{m}_{\hspace{1ex} j}})\sqrt{g} d^{4}x
\end{equation}
represents the effective Lagrangian after compactification.

\subsection{An explicit solution}\label{ExplicitConstantCurvature}

Using the result above Eq.(\ref{ABgen}) one can show that an
extension of the identifications that give rise to the BTZ black
hole \cite{Banados:1993gq} in higher dimensions leads to a space
divided in three regions with
\begin{equation}\label{AB}
A(r)^{2}= l^2 \left[\frac{r^2-r^2_{-}}{r^2_{+}-r^2_{-}}\right],
\qquad B(r)^{2}= l^{2}
\left[\frac{r^2-r^2_{+}}{r^2_{+}-r^2_{-}}\right]
\end{equation}
and metric defined by
\begin{description}
\item[region I] $r_{+}<r$
\begin{equation}\label{refionIme}
    ds_{I}^2 =  B(r)^2 d\Sigma_{1}^{L} + \left((B')^2 - (A')^2 \right) dr^2 + A(r)^{2}d\Sigma_{-1}^{E},
\end{equation}

\item[region II] $r_{-}<r<r_{+}$
\begin{equation}\label{refionIIme}
    ds_{II}^2 =  -B(r)^2 d\Sigma_{-1}^{E} + \left((B')^2 - (A')^2 \right) dr^2  + A(r)^{2}d\Sigma_{-1}^{E},
\end{equation}

\item[region III] $ r < r_{-}$
\begin{equation}\label{refionIIIme}
    ds_{III}^2 =  -B(r)^2 d\Sigma_{-1}^{E} + \left((B')^2 - (A')^2 \right) dr^2  + A(r)^{2}d\Sigma_{1}^{L},
\end{equation}
\end{description}
Here $d\Sigma$ stands for the line element of a submanifold. The
form of each of these (sub)manifolds is known but unnecessary for
this discussion . It is enough to know that the subindexes represent
the normalized curvature, and the indexes $L$ or $E$ stands for
Lorentzian or Euclidean manifold. Therefore for $r>r_{+}$ the
worldsheet is actually a cosmology, which for $r_{-}<r<r_{+}$
becomes an Euclidean hyperbolic space. At first sight  it seems that
jump occurs at $r=r_{+}$ , however the smooth vanishing of
$B^{2}(r)$ as $r\rightarrow r_{+}$ actually determines a smooth
change between positive and negative curvatures.

From Eq.(\ref{ABgen}) one can also construct a solution without
horizons which globally can be described by
\begin{equation}\label{SolHorizonless}
    ds_{I}^2 =  B(r)^2 d\Sigma_{-1}^{L} + \left(-(B')^2 + (A')^2 \right) dr^2 + A(r)^{2}d\Sigma_{1}^{E},
\end{equation}
where the constants in Eq.(\ref{ABgen}) are positive.

There is another solution, which can be casted as the
\textit{extreme} limit, $r_{-}\rightarrow r_{+}$, of the solution
(\ref{AB}). As for 2+1 dimensional black hole this solution also can
be obtained through an identification
\cite{Banados:1993gq,Steif:1996zm}. In this solution the worldsheet
and transverse sections are flat and the space is divided in two
regions described by the metrics,
\begin{description}
\item[region outside] $r_{+}<r$
\begin{equation}\label{refionImeExtreme}
    ds_{out}^2 =  B(r)^2 d\Sigma_{0}^{L} + D(r)^2 dr^2 + B(r)^{2}d\Sigma_{0}^{E},
\end{equation}

\item[region inside] $ r < r_{+}$
\begin{equation}\label{refionIIImeExtreme}
    ds_{in}^2 =  -B(r)^2 d\Sigma_{0}^{E} + D(r)^2 dr^2  - B(r)^{2}d\Sigma_{0}^{L},
\end{equation}
\end{description}
where
\[
B(r)^{2} = l^2\frac{r^2-r^2_{+}}{2r_{+}}\qquad D(r)^2 = \frac{
r^2}{(r^2-r^2_+)^2}
\]

\acknowledgments

We would like to thank Professors C. Martinez, R. Troncoso, J.
Zanelli for some very interesting and delighting discussions. R.A.
would like to thank Abdus Salam International Centre for Theoretical
Physics (ICTP) for its support. This work was partially funded by
grants FONDECYT 1040202 and DI 06-04. (UNAB).


\providecommand{\href}[2]{#2}\begingroup\raggedright\endgroup

\end{document}